\documentclass[aps,pre,
twocolumn,
superscriptaddress,%
centertags, showkeys]%
{revtex4-1}

\usepackage[utf8]{inputenc}
\usepackage{amsmath}
\usepackage{amsfonts}
\usepackage{amssymb}
\usepackage[all]{xy}
\usepackage{graphicx}
\usepackage{color}

\DeclareMathOperator{\arctanh}{arctanh}
\DeclareMathOperator{\arcsinh}{arcsinh}

\newcommand{\idots}{\text{\reflectbox{$\ddots$}}}

\newcommand{\cc}[2]{\underbrace{#2}_{c_{#1}}}

\begin{document}

\title{Random telegraph processes with non-local memory}

\author{S.S.~Apostolov}
\affiliation{O.~Ya.~Usikov Institute for Radiophysics and Electronics NASU, 61085 Kharkiv, Ukraine}

\author{O.V. Usatenko}
\affiliation{O.~Ya.~Usikov Institute for Radiophysics and Electronics NASU, 61085 Kharkiv, Ukraine}
\affiliation{Center for Nonlinear Science, University of North Texas, P.O. Box 311427, Denton, TX 76203-5370, USA}
\affiliation{Department of Physics, University of Florida, P. O. Box 118440, Gainesville, FL 32611-8440, USA}
\affiliation{Instituto de F\'{i}sica, Benem\'{e}rita Universidad Aut\'{o}noma de Puebla, Puebla, Pue. 72570, M\'{e}xico}

\author{V.~A.~Yampol'skii}
\affiliation{O.~Ya.~Usikov Institute for Radiophysics and Electronics NASU, 61085 Kharkiv, Ukraine}
\affiliation{V.~N.~Karazin Kharkiv National University, 61077 Kharkiv, Ukraine}

\author{S.~S.~Melnyk}
\affiliation{O.~Ya.~Usikov Institute for Radiophysics and Electronics NASU, 61085 Kharkiv, Ukraine}

\author{P.~Grigolini}
\affiliation{Center for Nonlinear Science, University of North Texas, P.O. Box 311427, Denton, TX 76203-5370, USA}

\author{A.~Krokhin}
\affiliation{Center for Nonlinear Science, University of North Texas, P.O. Box 311427, Denton, TX 76203-5370, USA}

\begin{abstract}
We study two-state (dichotomous, telegraph) random ergodic continuous-time processes with dynamics depending on their past.  We take into account the history of process in an explicit form by introducing an integral non-local memory term into the conditional probability function. We start from an expression for the conditional transition probability function describing additive multi-step binary random chain and show that the telegraph processes can be considered as continuous-time interpolations of discrete-time dichotomous random sequences. An equation involving the memory function and the two-point correlation function of the telegraph process is analytically obtained.
This integral equation defines the correlation properties of the processes with \emph{given memory functions}. It also serves as a tool for solving the inverse problem, namely for generation of a telegraph process with \emph{a prescribed pair correlation function}. We obtain analytically the correlation functions of the telegraph processes with two exactly solvable examples of memory functions and support these results by numerical simulations of the corresponding telegraph processes.
\end{abstract}


\maketitle


\section{Introduction}

The problems dealing with systems exhibiting long-range spatial and/or
temporal correlations remain to be on the top of
intensive research in physics, as well as in
theory of dynamical systems and in theory of probability~\cite{Bennett,Thurner,Ladyman,Rong,Lavazza,Illiashenko}.
Nature offers a large number of examples
of random processes. Moreover, they occur even more often than those with a
deterministic behavior. A systematic research of these processes is
necessary to describe a vast range of complex phenomena.

A need to generate a correlated random process of continuous or discrete variable appears in many areas of physics and engineering. The progress in this field of research may have a strong impact on design of a new class of electronic
nano-devices, optic fibers, acoustic and electromagnetic wave-guides
with selective transport properties (see, e.g.,
Refs.~\cite{Rice54,WOD95,IzKr99,IzMak0103,IKMrew}. The key
ingredient of the theory of correlated disorder is the two-point
(pair or binary) correlator of a random process. As was shown for a
weak disorder, this correlator fully determines the
transmission/reflection of classical or quantum waves through
disordered structures. The algorithm proposed in publications~\cite{Rice54,WOD95,IzKr99,IzMak0103,IKMrew}) generates a statistical ensemble of random functions (trajectories of the process) all possessing the same pair correlator. Generally, the random values of the functions are not limited and they may take any number from $-\infty$ to $\infty$. In this work, we study a wide class of processes when random variable takes only two values, say $a$ and $b$. Such processes are often found in nature; they are referred to as \emph{telegraph processes}, also known under the names Kac and dichotomous random processes.

The study of telegraph processes has a long history and is of grate interest to researchers. Thus, the classic of probability A. Kac writes in his work~\cite{Kac} ``We will consider a very simple stochastic model, a random walk. Unfortunately, this model is little known. It has very interesting features and leads not to a diffusion equation but to a hyperbolic one. The model first appeared in the literature in a paper by Sidney Goldstein, known to you mostly because of his work in fluid dynamics. The model had first been proposed by G. I. Taylor -- I think in an abortive, or at least not very successful, attempt to treat turbulent diffusion. But the model itself proved to be very interesting''. At present, the telegraph process has been studied to a much greater extent than at the time Kac's work was published. Currently application of the theory of random telegraph processes can be found in a variety of complex phenomena. To mention a few, ion channel gating dynamics in biological transport processes and gene expression levels in cells, motion of bacteria, neuronal spike trains, disorder-induced spatial patterns, first-passage and thermally activated escape processes, some aspects of spin dynamics, hypersensitive transport, stochastic resonance, quantum multifractality, blinking quantum dots,
rocking ratchets, and intermittent fluorescence. The diverse dichotomous systems may display non-ergodicity and/or L\'{e}vy statistics. Links and references to these and many other important studies related to numerous applications of dichotomous processes can be found in Refs.~\cite{Allegrini,Margolin,Sandev}.

The telegraph process is of interest not only from pure mathematical point of view, but also as a mechanism of specific noise affecting some dynamical systems. If the noise is neither Gaussian nor dichotomous, then it is generally impossible to analyze its effect on an dynamical system.

One of the ways to study the nature of correlations in a dynamical system is to model the system evolution by a mathematical object (for example, a correlated sequence of symbols) possessing the same statistical properties as the system itself. Several algorithms for generation of random sequences with long-range correlations are known in  literature~\cite{Rice54,WOD95,IzKr99,IzMak0103,IKMrew,Stanley,Czirok,Izrailev}.
Here we propose a powerful method based on the statistics of multi-step Markov chains. The additive Markov chain models~\cite{UYa,RewUAMM,MUYa} have shown their effectiveness in describing diverse objects, including literary texts and DNA sequences; therefore, it is of undoubted interest to obtain a generalization of these models to the class of systems characterized by continuous parameters.

The Markov process is a common and natural tool for describing a random phenomena (see, e.g.,~\cite{zab,reb,fer,nic}). Two well-known Gaussian
Markov processes -- Brownian motion and Ornstein-Uhlenbeck
process~\cite{Uhlen} -- have been used extensively in various
applications from financial mathematics to natural sciences~\cite{Kampen,gar,Horsthemke}. Both these processes can be described by the Langevin equation~\cite{gar} for the random variable $V(t)$ (e.g., for the velocity of the particle),
\begin{eqnarray} \label{-2}
dV(t)=-\nu V(t) dt  + \sigma \, dW(t).
\end{eqnarray}
Here $dW(t)$ is the standard centered white noise. The term $-\nu V(t) dt$  describes a linear friction between the particle and the bath.  It is important to note that such an equation is valid only if the external random applied force is a Gaussian white noise. In this case, the friction force is a linear function of the random variable $V(t)$. In a more general case, the friction force is a linear functional depending on the entire past dynamics of the system and can be written in the form,
\begin{equation} \label{12}
d V(t)_{mem}=\left( \int_{0} ^{\infty} \mu(t')
V(t-t')dt'\right)dt,
\end{equation}
(see Refs.~\cite{Zwanzig60,Mori,Zwanzig,Ford}). Thus, Eq.~\eqref{-2} with additional term Eq.~\eqref{12} containing the memory kernel $\mu(t')$  becomes an integro-differential equation and describes a non-Markov process. By definition, all non-markovian processes are history-dependent.

We may not know the nature and statistical characteristics of the random forces applied to the system, but it follows from the example of Langevin equation that if the applied force is not a delta-correlated process then the additional terms should appear in Eq.~\eqref{-2}, e.g., in the form of Eq.~\eqref{12}.

In this paper, we take into account the history of telegraph process in the explicit form introducing an integral non-local memory term into the transition conditional probability function. A telegraph process with memory can be used to describe a wider range of phenomena than an ordinary telegraph process without memory.

The relation between the correlation and memory functions is a rather complex integro-differential equation which cannot be solved in general case. Here we demonstrate two interesting particular cases when the solution can be obtained analytically.  Note an important point. The equation for the telegraphic process with the added memory term considered here does not describe a renewal process~\cite{Cox}.

The structure of the paper is as follows.
In Section \ref{GenDef}, we present some general definitions and provide a brief description of the models and the relevant previous results. We start from an expression for transition conditional probability function describing additive multi-step random chain and show that the proposed processes can be considered as generalization to continuous variable of a discrete-time random markovian sequence. In addition, an equation connecting the memory function and the two-point correlation function of the process is obtained.
In Section~\ref{Special}, we solve analytically the equations for correlation function for some special examples of the memory function. The last Section~\ref{Discussion} contains conclusions and the outline for further research.

\section{Telegraph process with memory as a generalization of the discrete multi-step Markov chain} \label{GenDef}
 A random process $N(t)$ that represents the total number of occurrences of an event within the time interval (0, $t$] is called a renewal process, if the time intervals between failures are independent and identically distributed random variables. The Poisson and telegraph  processes with exponentially distributed intervals between events are the well-known particular cases of the renewal processes.

In the conventional probability theory, a telegraph process is a memoryless continuous-time stochastic process where the random variable can take on two distinct values only, say $a$ and $b$. It describes, for example, a one-dimensional random motion of a particle moving with a constant velocity $v=a$ along some direction for some random time interval drawn
from an exponential distribution, and after that, the particle moves to the opposite direction with the velocity $b$, where $b=-a=-v$.
Thus, we declare that, independently of prehistory of the particle motion, the probabilities to generate the random value of $x_{t+dt}$ are:
\begin{subequations}
\label{prob-tel-no-mem}
\begin{eqnarray}\label{prob-tel-no-mem1}
P(x_{t+dt}=b|x_{t}=a) =  \lambda dt,
\\[6pt] P(x_{t+dt}=a|x_{t}=b) = \mu dt,
\end{eqnarray}
where the random process is defined by two  constants, $\lambda$ and $\mu$, representing the inverse average times of life $1/\overline{t}_a $ and $1/ \overline{t}_b$ of the particle in the states $a$ and $b$, correspondingly. The counterparts of these equations are the following relations:
\begin{eqnarray}\label{prob-tel-no-mem2}
P(x_{t+dt}=a|x_{t}=a) = 1 - \lambda dt,
\\[6pt]
P(x_{t+dt}=b|x_{t}=b) = 1 - \mu dt.
\end{eqnarray}
\end{subequations}

\begin{figure}[h!]
\center\includegraphics[width=0.53\textwidth]{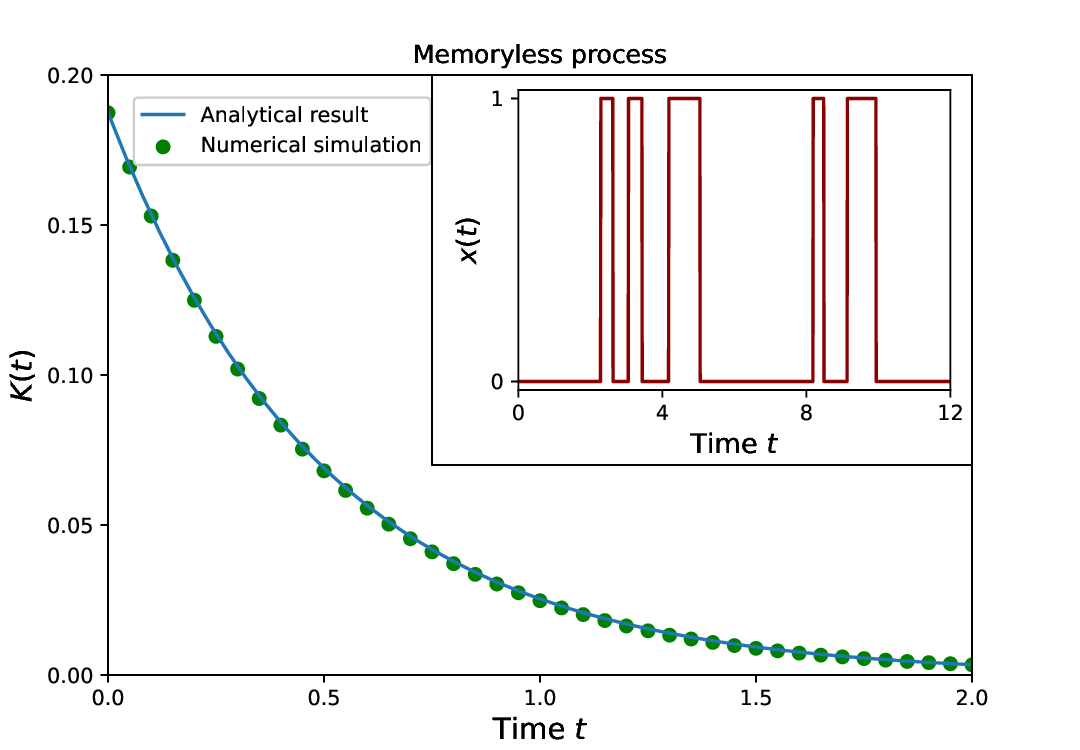}
\caption{(Color online) The telegraph process without
memory.  The correlation function, $K(t)=\mu\lambda/(\mu+\lambda)^2\exp [-(\mu+\lambda) t]$, is plotted analytically (continuous curve) and numerically (dots). The parameters of the generated process are $\lambda = 1.5, \mu = 0.5$, $a=1$,
$b=0, \overline{x} = 0.25$. The length of process time is $10^5$ and the time step of generation is $10^{-2}$. The insert shows the generated random variable $x(t)$.} \label{Fig1}
\end{figure}

If the life time of the system (without memory) in the states $a$ and $b$ is governed by Eqs.~\eqref{prob-tel-no-mem}, then it is possible to construct the process by two methods: step-by-step generation with infinitesimally small time-step $dt$, or global generation of random time intervals $t_a$ and $t_b$ of the system to stay in the states $a$ and $b$. These two ways are equivalent for the processes without memory. However, the first method allows to adequately include memory into the process. Therefore, we use namely this method in our numerical simulations. A fragment of numerically constructed telegraph process without memory is shown in the insert to Fig.~\ref{Fig1}.

The conditional probabilities, Eqs.~\eqref{prob-tel-no-mem}, ``know'' only the current values of the random variable, i.e., taken at instant $t'=t$. To take into account the memory effects 
from previous times $t'<t$, we have to introduce integral memory in the form similar to Eq.~\eqref{12} (see also Eqs.~\eqref{prob-tel-mem}) to the right-hand sides of  Eqs.~\eqref{prob-tel-no-mem}. However, we  prefer here a more transparent and clear way by considering the analogy of the telegraph process with a discrete additive memory-dependent Markov chain. The following step is a transition from the discrete random sequence to the continuous-time random process.

A convenient representation of a discrete random chain is to write down its transition conditional probability function \cite{UYa,RewUAMM,MUYa}. This function completely determines the dynamics of the random chain as well as its correlation properties. The transition conditional probability function $P(x_{r+1}=\alpha|x_{r+1-N},\ldots,x_{r-1},x_{r})$  of the binary, $\alpha=\{0,1\}$,
$N$-step Markov chain is written as follows \cite{MUYa},
\begin{equation}\label{CondPr_power}
  P\left( x_{r+1}=1 | x_{r+1-N}^{r} \right)
    =  \overline{x} + \sum_{r'=1}^{N} F(r') (x_{r+1-r'}-\overline{x} ),
\end{equation}
where $\overline{x}$ is the average value of the random variable $x$ and the concise notation $x_{r+1-N}^{r}=x_{r+1-N},...,x_{r}$ for a sequence of $N$ previous random values is used.

There is no admitted name for the random sequences defined by Eq.~\eqref{CondPr_power}. It can be referred to as categorical~\cite{Hoss},
higher-order~\cite{Raftery, Seifert}, multi- or
$N$-step~\cite{UYa,RewUAMM} Markov's chains. One of the most
important and interesting application of the symbolic sequences is
the probabilistic language model, which specializes in predicting
the next item in a sequence by means of $N$ previous known symbols.
In this sense the Markov chains are known as the $N$-gram models.
 We refer to such sequences as the additive Markov chains and $F(r)$
as the \emph{memory function}. It describes the strength of
influence of the previous symbols $x_{r+1-r'} \,\,\,(1 \leqslant r' \leqslant N$)
upon a generated one, $x_{r+1}$.

Let us rewrite Eq.~\eqref{CondPr_power} for the conditional probability function of the binary Markov chain of $\{a,b\} =\{1,0\}$ in the equivalent form,
\begin{eqnarray} \label{Before}
&P&(x_{r+1}=1|x_{r}; x_{r'<r}) = \bar{x} +(1-2\gamma \Delta t)(x_{r}-\bar{x})
\nonumber\\[6pt] &+& \sum_{r'=1}^{\infty}\alpha_{r'}(\Delta t)^2(x_{r-r'}-\bar{x}),
\end{eqnarray}
with
\begin{equation} \label{AvX}
\bar{x}=\mu/(\mu+\lambda), \quad\gamma=(\mu+\lambda)/2.
\end{equation}
The  term proportional to $\Delta t$ describes the influence of the nearest term $x_{r}$ on the generated symbol $x_{r+1}$ and the terms proportional to $(\Delta t)^2$ are converted to the integral memory contributions in the limit $\Delta t \rightarrow 0$.

The random sequence defined by Eq.~\eqref{Before} is stationary since the conditional probability function $P(x_{r+1}=1|x_{r}; x_{r'<r})$ does not depend \emph{explicitly} on the discrete coordinate $r$. The sequence is ergodic if the conditional probability function satisfies the \emph{strict} inequalities,
\begin{equation} \label{BeforeErgod}
0<P(x_{r+1}=1|x_{r}; x_{r'<r}) < 1,
\end{equation}
that impose certain restrictions on the sequence parameters $\lambda, \mu$ and  the function $ \alpha_r$.

Transformation to the continuous time in Eq.~\eqref{Before} occurs in the limit $r\rightarrow \infty, \Delta t \rightarrow 0$ and
\begin{equation}
\label{cont-trans}
r\Delta t \to t,
\quad
r'\Delta t \to \tau,
\quad
\Delta t \to dt,
\quad
\alpha_{r'}\to\alpha(\tau).
\end{equation}
This transformation leads to \textit{the telegraph process with memory} where the conditional probabilities are given by:

\begin{widetext}
\begin{subequations}
\label{prob-tel-mem}
\begin{eqnarray}\label{cont-trans0}
P(x_{t+dt}=1|x_{t}=1; x_{t'<t}) &=& 1 - \big[\lambda-\int_0^\infty \alpha(\tau)(x_{t-\tau}-\bar{x})d\tau\big] dt,
\\
P(x_{t+dt}=0|x_{t}=1; x_{t'<t}) &=&  \big[\lambda-\int_0^\infty \alpha(\tau)(x_{t-\tau}-\bar{x})d\tau\big] dt,
\\
P(x_{t+dt}=0|x_{t}=0; x_{t'<t}) &=& 1 - \big[\mu+\int_0^\infty \alpha(\tau)(x_{t-\tau}-\bar{x})d\tau\big] dt,
\\
P(x_{t+dt}=1|x_{t}=0; x_{t'<t}) &=& \big[\mu+\int_0^\infty \alpha(\tau)(x_{t-\tau}-\bar{x})d\tau\big] dt.
\end{eqnarray}
\end{subequations}
\end{widetext}
These equations are the generalization of basic definitions
\eqref{prob-tel-no-mem}. In what follows, we will call $\alpha(\tau)$ as the memory function of the telegraph process. The integral terms in Eqs.~\eqref{prob-tel-mem} describe a memory effect on the process but do not change the average value $\bar{x}$, since the integral term averages are zero.

The important statistical characteristics of a random process is the correlation function. In order to get the relation between the memory and correlation functions we start from obtaining the similar equation for the random discrete Markov chains with memory using the well known definition of the correlation function,
\begin{equation}\label{DefK}
K_{r}=\overline{(x_{i+r}-\bar{x})(x_{i}-\bar{x})}=\overline{x_{i+r}x_{i}}-\bar{x}^2.
\end{equation}
Multiplying Eq.~\eqref{Before} by $x_{0}$ and averaging over the ensemble of random sequences, we derive an equation for the correlation function of the random sequence,
\begin{equation}\label{EqK}
K_{r+1}=(1-2\gamma\Delta t) K_r+\sum_{r'=1}^{\infty}\alpha_{r'}(\Delta t)^2K_{r-r'}, \qquad r>0.
\end{equation}
This relation can be obtained also by averaging over the coordinate $r$ along the chain (see Ref.~\cite{MUYaAM06}). The coincidence of results of these two methods of averaging follows from the ergodicity of the sequences under study.

Note that the similar equation for correlation function is valid also for the autoregressive random sequences (see Yule-Walker equations in Refs.~\cite{MYaU,Yule,Walker}).

Rewriting Eq.~\eqref{EqK} in the following form,
\begin{equation}
\label{markovK}
\dfrac{K_{r+1}-K_r}{\Delta t}=-2\gamma K_r+\sum_{r'=1}^{\infty}\alpha_{r'}\Delta tK_{r-r'}, \qquad r>0,
\end{equation}
and taking limit~\eqref{cont-trans} we obtain the integro-differential equation for the correlation function $K_r\to K(t)$ of the random telegraph process,
\begin{equation}
\label{forK}
\frac{dK(t)}{dt}+2\gamma K(t)=\int_0^\infty\alpha(\tau)K(t-\tau)d\tau,
\qquad
t>0.
\end{equation}

The solution of Eq.~\eqref{forK} is subject to the initial condition,
\begin{equation}
\label{init}
K(0)=\dfrac{\mu\lambda}{(\mu+\lambda)^2},
\end{equation}
and, according to definition, the parity of correlation function,
\begin{equation}
\label{even}
K(-t)=K(t),
\qquad
t>0.
\end{equation}

The exponential solution $K(t)=\mu\lambda/(\mu+\lambda)^2\exp [-(\mu+\lambda) t]$ of Eq.~\eqref{forK} with conditions Eqs.~\eqref{init} -~\eqref{even} at $\lambda = 1.5, \mu = 0.5$ and $\alpha(\tau) = 0$ is presented by solid line in the main panel in Fig.~\ref{Fig1}. The filled circles on this curve show the results of numerical simulation of the correlation function $K(t)$ for the memoryless telegraph process presented in the insert.

Note that, in principle, the term proportional to $\gamma$ in Eq.~\eqref{forK} can be included into the integral term by adding the appropriate delta-function to the memory function $\alpha (\tau)$.

It is necessary to emphasize that relation~\eqref{forK} can be used also for solving an inverse problem of finding the unknown memory function $\alpha(t)$ when the correlation function $K(t)$ is given. Once $\alpha(t)$ is calculated, we can generate the telegraph process with arbitrary prescribed correlation function using the transition conditional probability functions Eqs.~\eqref{prob-tel-mem} as it was done in Ref.~\cite{MUYa} for random additive multi-step Markov sequences.

\section{Special cases for memory}\label{Special}


The integral term in Eq.~\eqref{forK} does not have a form of convolution, therefore this equation with conditions~\eqref{init} and~\eqref{even} cannot be solved, in general case, applying the Fourier or Laplace transform. In this section we find the solutions for two particular forms of the memory function~$\alpha(\tau)$.

\subsection{$\delta$-delayed memory}

We start from the case of memory function,
\begin{equation}
\alpha(\tau)=\zeta\delta(\tau-T),
\end{equation}
which takes into account the memory of the process at only one point of the past at $t=T$. Then the conditional probabilities Eqs.~\eqref{prob-tel-mem} are rewritten as
\begin{widetext}
\begin{subequations}
\begin{eqnarray}
P(x_{t+dt}=1|x_{t}=1; x_{t'<t}) &=& 1 - \big[\lambda-\zeta(x_{t-T}-\bar{x})\big] dt,
\\[6pt]
P(x_{t+dt}=0|x_{t}=1; x_{t'<t}) &=&  \big[\lambda-\zeta(x_{t-T}-\bar{x})\big] dt,
\\[6pt]
P(x_{t+dt}=0|x_{t}=0; x_{t'<t}) &=& 1 - \big[\mu+\zeta(x_{t-T}-\bar{x})\big] dt,
\\[6pt]
P(x_{t+dt}=1|x_{t}=0; x_{t'<t}) &=& \big[\mu+\zeta(x_{t-T}-\bar{x})\big] dt.
\end{eqnarray}
\end{subequations}
\end{widetext}
It should be noted that the possible values of parameter $\zeta$ are constrained by conditions
\begin{equation}
- \rm{min}\left(\frac{\lambda}{\mu}, \frac{\mu}{\lambda}\right)(\lambda+\mu)<\zeta<\lambda+\mu,
\end{equation}
which guarantee the natural property of probability, $0<P(...)<1$.

Such a memory yields the following delay differential equation for the correlation function:
\begin{equation}
\label{forK-delay}
\frac{dK(t)}{dt}+2\gamma K(t)=\zeta K(t-T),
\qquad
t>0.
\end{equation}
We solve this equation in three steps:

1. Find the solution~$K(t)=K_0(t)$ for $0<t<T$ with
\begin{eqnarray}
\label{init-func}
K_0(t)&=&K(0)\dfrac{\cosh(\phi_0-\eta t)}{\cosh\phi_0},
\\[6pt]
\phi_0&=&\dfrac{\eta T}{2}+\arctanh\dfrac{2\gamma-\zeta}{\eta},  \eta =\sqrt{4\gamma^2-\zeta^2}>0.\nonumber
\end{eqnarray}

2. Present the solutions for the time intervals, $nT<t<(n+1)T$, in the following form,
\begin{eqnarray} \label{eq:Kn-delta}
K_n(t)&=&K(0)\dfrac{\cosh[\phi_n-\eta (t-nT)]}{\cosh\phi_0} \\[6pt]
&+&P_n(t)\exp[-2\gamma(t-nT)],\nonumber
\end{eqnarray}
where $P_n(t)$ are $(n-1)$-th degree polynomials, and
\begin{eqnarray}
\phi_n=\phi_0+n\arctanh\frac{\eta }{2 \gamma }.\nonumber
\end{eqnarray}

3. Obtain the recurrence relation for the polynomials~$P_n(t)$,
\begin{eqnarray}\label{P-diffeq}
\frac{dP_n(t)}{dt}=\zeta P_{n-1}(t-T),
\end{eqnarray}
and the continuity condition for the correlation function $K(t)$ at $t=nT$,
\begin{eqnarray}\label{P-bound}
&&P_n(nT)=
P_{n-1}(nT)\exp(-2\gamma T)\\[6pt]&&+\dfrac{K(0)}{\cosh\phi_0}\big[\cosh(\eta T-\phi_{n-1})-\cosh \phi_n\big].\nonumber
\end{eqnarray}
%
and analyze them.  The mathematical details of calculations leading to the explicit  form for $K(t)$ are presented in Appendix~\ref{DeltaMem}. The final results for the correlation functions $ K_1(t)$ and $ K_2(t)$ are following:
\begin{eqnarray}\label{K-Delta1}
       K_1(t)&=&K(0)\dfrac{\cosh[\phi_1-\eta (t-T)]}{\cosh\phi_0}\\ [6pt] &+&A_1\exp[-2\gamma(t-T)], \quad T<t<2T, \nonumber
\end{eqnarray}
\begin{eqnarray}\label{K-Delta2} K_2(t)&=&K(0)\dfrac{\cosh[\phi_2-\eta (t-2T)]}{\cosh\phi_0}\\ [6pt] &+&[\zeta A_1 (t-2T)+A_1\exp(-2\gamma T) + A_2]\nonumber\\ [6pt]
  &\times&\exp[-2\gamma(t-2T)],\,\, 2T<t<3T,\nonumber
\end{eqnarray}
with
\begin{eqnarray}\label{init-func-last}
    A_1&=&\dfrac{K(0)}{\cosh\phi_0}\big[\cosh(\eta T-\phi_{0})-\cosh \phi_1\big],\\[6pt]
    A_2&=&\dfrac{K(0)}{\cosh\phi_0}\big[\cosh(\eta T-\phi_{1})-\cosh \phi_2\big]. \nonumber 
\end{eqnarray}

\begin{figure}[h!]
\begin{centering}
\scalebox{0.46}[0.57]{\includegraphics{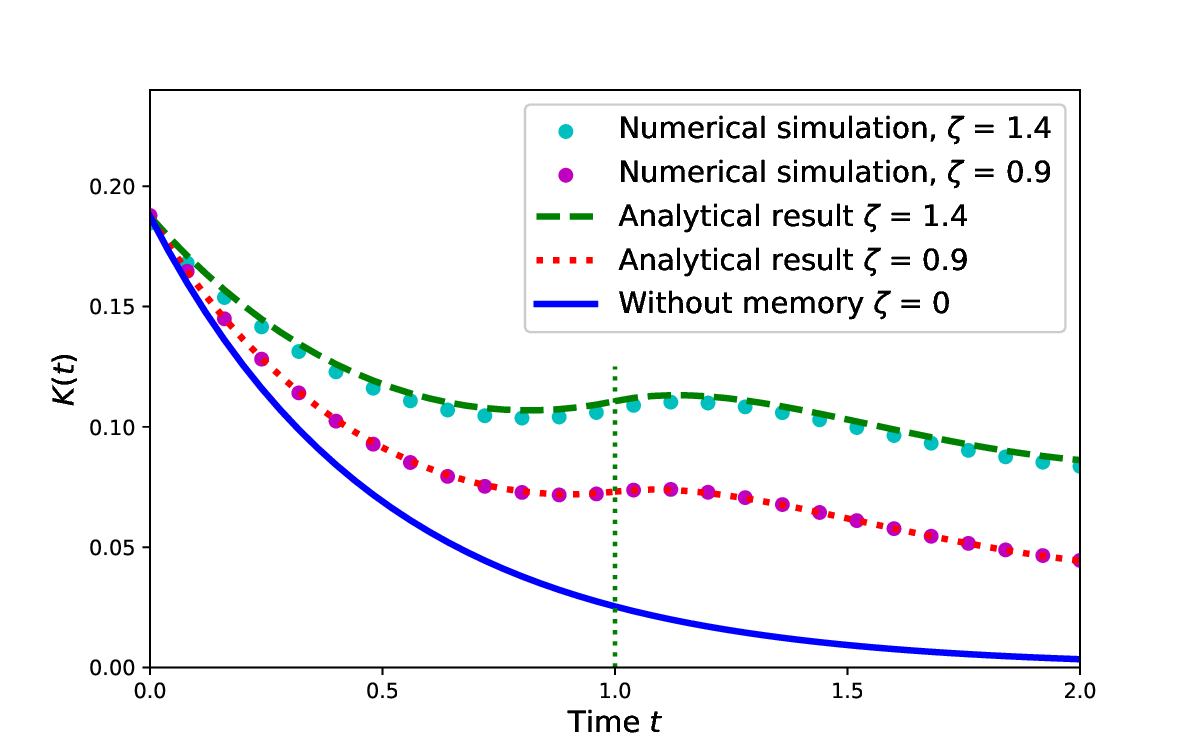}}
    \caption{(Color online) The correlation functions of the telegraph processes with delta-delayed memory (dashed and dotted lines with symbols) and without memory (solid line). The parameters of generated processes are the same as in Fig.~\ref{Fig1}. The dashed and dotted lines present the correlation functions given by Eqs.~\eqref{init-func} and~\eqref{K-Delta1}, the symbols are the results of numerical simulations.  The values of memory parameter $\zeta$ are shown in the legend. The vertical line at $t = T = 1$ indicates the singular point position for memory-dependent processes.}
\label{Fig2}
\end{centering}
\end{figure}
Correlation functions $K_0(t)$ and $K_1(t)$ of the process with different values of memory constant $\zeta$ are presented in Fig.~\ref{Fig2}. Let us pay attention to the specific property of the correlation function $K(t)$ of the process with $\delta$-delayed memory. The function $K(t)$ being itself continuous has a discontinuity of its $(n+1)$-th derivative at $t=n T, \,\,n = 0,1,2,...$  Indeed, one can see in Fig.~\ref{Fig2} that $K'(t)$ is discontinuous at $t=0$ (recall that $K(-t)=K(t)$) and $K''(t)$ is discontinuous at $t=T=1$.
\subsection{Step-wise memory}

In this subsection, we study the telegraph process with the step-wise memory function,
\begin{eqnarray}\label{StepAlpha}
\alpha(\tau)=\xi[\theta(\tau)-\theta(\tau-T)],
\end{eqnarray}
where $\theta(.)$ is the Heaviside step function. In this case, the transition conditional probability functions can be written in the form:
\begin{widetext}
\begin{subequations}\label{Step-wise}
\begin{eqnarray}
P(x_{t+dt}=1|x_{t}=1; x_{t'<t}) &=& 1 - \Big[\lambda-\xi\int_0^T(x_{t-\tau}-\bar{x})d\tau\Big] dt,
\\[6pt]
P(x_{t+dt}=0|x_{t}=1; x_{t'<t}) &=&  \Big[\lambda-\xi\int_0^T(x_{t-\tau}-\bar{x})d\tau\Big] dt,
\\[6pt]
P(x_{t+dt}=0|x_{t}=0; x_{t'<t}) &=& 1 - \Big[\mu+\xi\int_0^T(x_{t-\tau}-\bar{x})d\tau\Big] dt,
\\[6pt]
P(x_{t+dt}=1|x_{t}=0; x_{t'<t}) &=& \Big[\mu+\xi\int_0^T(x_{t-\tau}-\bar{x})d\tau\Big] dt.
\end{eqnarray}
\end{subequations}
\end{widetext}
The possible values of the parameter $\xi$ are constrained by the conditions
\begin{eqnarray}\label{StepXi}
- {\rm min}\left(\frac{\lambda}{\mu}, \frac{\mu}{\lambda}\right)(\lambda+\mu)< \xi T < \lambda+\mu.
\end{eqnarray}

From Eq.~\eqref{forK}, we obtain the following integro-differential equation for the correlation function of telegraph process with the step-wise memory function:
\begin{eqnarray}
\label{forK-step}
\frac{dK(t)}{dt}+2\gamma K(t)=\xi\int_0^TK(t-\tau)d\tau,
\qquad t>0.
\end{eqnarray}
The detailed solution of this equation is given in Appendix~\ref{StepMemA}.
Here we present the result for $K(t)$ in two first time intervals:
\begin{eqnarray} \label{StepMemCF-0}
K(t) = K(0)\frac{2 \xi  \cosh \phi_0+\eta  (2 \gamma -\xi  T) \sinh (\phi_0-\eta  t)}{2 \xi  \cosh \phi_0+\eta (2 \gamma -\xi  T)  \sinh \phi_0 }
\end{eqnarray}
for  $0<t<T$ and
\begin{eqnarray} \label{StepMemCF}
&&K(t)=K(0)\frac{2 \xi  \cosh \phi_0-\eta  (2 \gamma -\xi  T) \sinh [\phi_1-\eta  (t-T)]}{2 \xi  \cosh \phi_0+\eta (2 \gamma -\xi  T)  \sinh \phi_0 }\nonumber \\[6pt]
&&+ 4 K(0)\frac{\cosh \phi_0 \exp{[-\gamma  (t-T)]}  (2 \gamma -\xi  T)(2 \gamma ^2+\xi) }{\xi \big[2 \xi  \cosh \phi_0+\eta (2 \gamma -\xi  T)  \sinh \phi_0\big] } \nonumber\\[6pt]
&&\times\Big\{2 \gamma  \cosh [\kappa  (t-T)]-\frac{2 \gamma ^2+\xi}{\kappa }\sinh [\kappa  (t-T)]\Big\}
\end{eqnarray}
for  $T<t<2T$.
Here
\begin{eqnarray}\label{phi01}
&&\eta =\sqrt{4\gamma^2+2\xi}>0, \quad \kappa=\sqrt{\gamma^2+\xi}, \\[6pt]
&&\phi_0=\frac{\eta  T}{2}+\arctanh\frac{2 \gamma }{\eta }, \quad \phi_1=\phi_0+\arcsinh\frac{2 \gamma  \eta }{\xi }.\nonumber
\end{eqnarray}

\begin{figure}[ht!]
\scalebox{0.46}[0.57]{\includegraphics{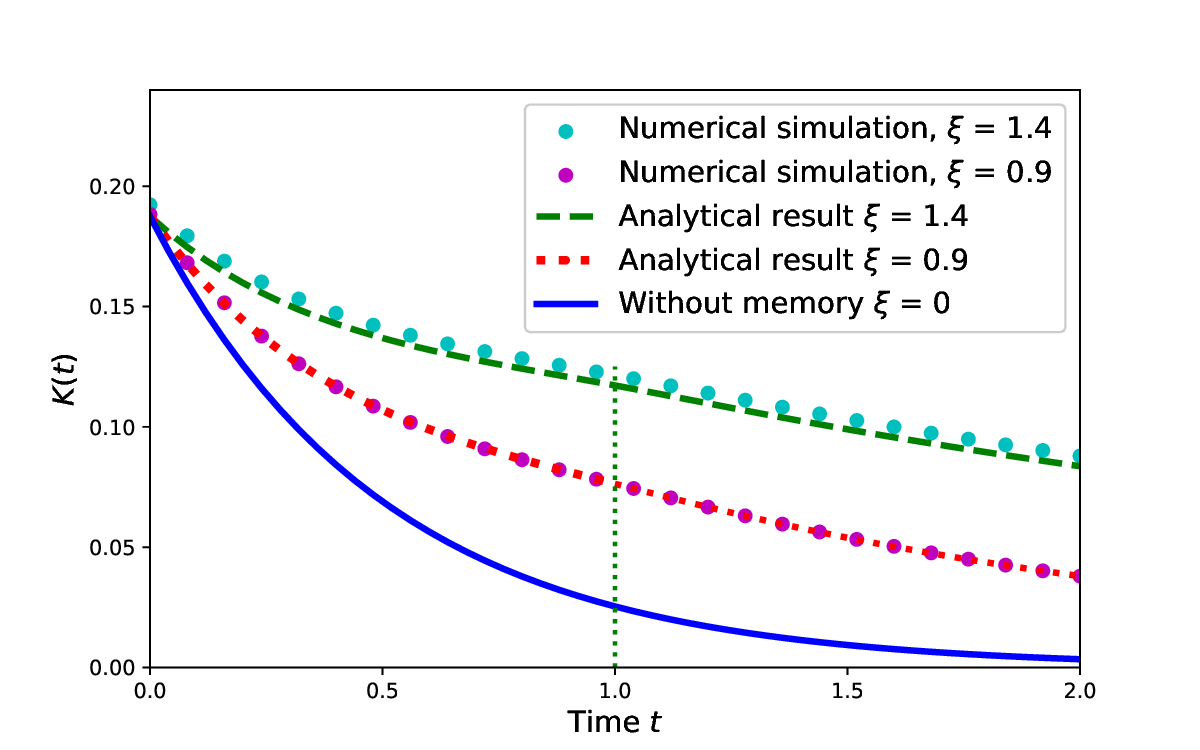}}
\caption{(Color online) The correlation functions of the telegraph processes with step-wise memory (dashed and dotted lines with symbols) and without memory (solid curve). The parameters of generated processes are the same as in Fig.~\ref{Fig1}. The dashed and dotted lines present the correlation functions given by Eqs.~\eqref{StepMemCF-0} and \eqref{StepMemCF}, the symbols are the results of numerical simulations.   The values of memory parameter $\xi$ are shown in the legend. The vertical line indicates the singular point position, $t = T = 1$, for memory-dependent processes.}  \label{Fig3}
\end{figure}

Correlation function $K(t)$ of the process with different values of memory constant $\xi$ is shown in Fig.~\ref{Fig3}. It is seen from Figs.~\ref{Fig2} and~\ref{Fig3} that the correlation functions obtained from the
numerically generated telegraph processes with different memory functions $\alpha (t)$ are in excellent
agreement with the corresponding results for $K(t)$ calculated analytically.

\section{Conclusion }\label{Discussion}

In conclusion, we propose a mathematical approach based on additive Markov chain to
study telegraph random ergodic processes with dynamics depending on the past. We took into account the history of the process in the explicit form introducing an integral non-local memory term into conditional probability function. We showed that the proposed processes can be considered as continuous-time interpolations of discrete-time higher-order random sequences. An equation connecting the memory function and the two-point correlation function of the telegraph process is obtained. This equation allows one not only to define the correlation properties of
processes with given memory function but can also serve as an instrument for solving the inverse problem of construction the telegraph process with a prescribed pair correlation function. We found analytically solutions of integral equations for the correlation functions of telegraph processes with delta-delayed and step-wise memory functions. As an illustration, some examples of numerical simulation of the processes with nonlocal memory are presented.

Natural continuation of this study is expansion of the proposed method to the processes with
time-dependent quantities $\lambda$ and $\mu$. This will allow, in particular, consideration of the effects of memory on telegraph processes with L\'{e}vy distributions of system life times in states $a$ and $b$ (see, e.g., Refs.~\cite{Bologna,Bologna2}). An interesting and separate problem is finding applications of the telegraph process with memory to specific random processes. In particular, the telegraph process can describe the information transcription in DNA molecules~\cite{Latchman} where the memory effects play extremely important role.

\begin{acknowledgments}
The authors thank D.~Maslov and P.~Hirschfeld for useful discussions and H.~Khajoei for assistance in numerical calculations. This work was partially supported by the EFRI grant no. 1741677 from the U.S. National Science Foundation. OU is grateful to H.-P.~Cheng for financial support and the Departments of Physics at the University of Florida and the University of North Texas for hospitality.
\end{acknowledgments}

\appendix 
\section{Solution of equation~\eqref{forK-delay} for the processes with $\delta$-correlated memory}\label{DeltaMem}
\textbf{Solution for $K_0(t)$.} Recalling parity condition~\eqref{even} we replace $K(t-T)$ by $K(T-t)$ for interval $0<t<T$ and rewrite Eq.~\eqref{forK-delay} as
\begin{eqnarray}
\label{forK-delay-0}
K_0'(t)+2\gamma K_0(t)=\zeta K_0(T-t).
\end{eqnarray}
Applying operator $d/dt-2\gamma$ to both sides of equation we get
\begin{subequations}
\begin{eqnarray}
&&K_0''(t)-4\gamma^2 K_0(t) = \zeta^2 K_0(t).
\end{eqnarray}
Its solution is
\begin{eqnarray}
&&K_0(t)=C_+ \exp(\eta t)+C_- \exp(-\eta t),
\\ [6pt]
&&\eta =\sqrt{4\gamma^2-\zeta^2}>0.\nonumber
\end{eqnarray}
Substituting this solution into Eq.~\eqref{forK-delay-0}, equating the coefficients at the exponents, after some algebra the solution is obtained in the form of Eq.~\eqref{init-func}.
\end{subequations}

\textbf{General solution by iterating procedure.} Let us denote $K(t)=K_n(t)$ for the interval $nT<t<(n+1)T$. Then Eq.~\eqref{forK-delay} naturally transforms in a sort of recurrence relation for functions~$K_n(t)$,
\begin{eqnarray}\label{Kn-diffeq}
&&K_n'(t)+2\gamma K_n(t)=\zeta K_{n-1}(t-T),\\ [6pt]
&&nT<t<(n+1)T, \nonumber
\end{eqnarray}
with boundary conditions $K_n(nT)=K_{n-1}(nT)$.

Then we solve the problem iteratively:

1. $K_0(t)$ is defined in Eq.~\eqref{init-func}.

2. Find $K_1(t)$ from Eq.~\eqref{Kn-diffeq} in the form of superposition of exponential functions $\exp(\pm \eta t)$ (particular solution sourced from hyperbolic cosine in $K_0(t)$ in the rhs) and $\exp(-2\gamma t)$ (general solution of homogeneous equation).

3. Find $K_2(t)$ from Eq.~\eqref{Kn-diffeq} in the form of superposition of exponential functions $\exp(\pm \eta t)$, $\exp(-2\gamma t)$ and $t\exp(-2\gamma t)$ (particular solution sourced from $\exp(-2\gamma t)$ in $K_1(t)$ in the rhs).

Continuing this procedure we can see from the procedure that the function~$K_n(t)$ can be written in the following form,
\begin{eqnarray}
&&K_n(t)=C_n\cosh[\eta (t-nT)-\phi_n]\\ [6pt]&&+P_n(t)\exp[-2\gamma(t-nT)],\nonumber
\end{eqnarray}
where $C_n$ and $\phi_n$ are constants and $P_n(t)$ is the polynomial function.

Substituting the last expression into Eq.~\eqref{Kn-diffeq} and equating prefactors of $\exp(\pm \eta t)$ and $\exp(-2\gamma t)$, we obtain the coefficients~$C_n$ and the recurrence relation for~$\phi_n$,
\begin{eqnarray}
C_n=\dfrac{K(0)}{\cosh\phi_0},\,\,\,\,\,\,
\phi_n=\phi_{n-1}+\arctanh \frac{\eta }{2 \gamma },
\end{eqnarray}
as well as the recurrence relation~\eqref{P-diffeq} with the continuity  condition~\eqref{P-bound} for the polynomials~$P_n(t)$.

\textbf{Iterative scheme for the polynomials~$P_n(t)$.} Proceeding iteratively, one can see that $P_{n}(t)$ is the polynomial of $(n-1)$-th degree. Then we can seek its explicit form as
\begin{eqnarray}
P_{n}(t)=\sum_{m=0}^{n-1} c_{nm}(t-nT)^m.
\end{eqnarray}

According to the recurrence relation~\eqref{P-diffeq}, we have
\begin{eqnarray}
&&\sum_{m=0}^{n-1} c_{nm}m(t-nT)^{m-1} \\
&&=\zeta\sum_{m=0}^{n-2} c_{(n-1)m}[t-T-(n-1)T]^m. \nonumber
\end{eqnarray}
Then coefficients~$c_{nm}$ with $m>0$ can be expressed via coefficients~$c_{(n-1)(m-1)}$, and, iteratively, via $c_{(n-m)0}$:
\begin{eqnarray}
&&c_{nm}=\dfrac{\zeta}{m}c_{(n-1)(m-1)}=\dfrac{\zeta^2}{m(m-1)}c_{(n-2)(m-2)}\\
&&=\ldots=\dfrac{\zeta^m}{m!}c_{(n-m)0}. \nonumber
\end{eqnarray}

Using the last relation and condition~\eqref{P-bound}, we obtain the recurrence relation for $c_{n0}$,
\begin{eqnarray}
c_{n0}= \sum_{m=0}^{n-2} \dfrac{(\zeta T)^m}{m!}c_{(n-m-1)0} \exp(-2\gamma T)+A_n,\nonumber
\end{eqnarray}
where
\begin{eqnarray}
A_n=\dfrac{K(0)}{\cosh\phi_0}\big[\cosh(\eta T-\phi_{n-1})-\cosh \phi_n\big].\nonumber
\end{eqnarray}

The scheme to calculate the coefficients $c_{nm}$ is presented in the following diagram:
\begin{equation*}
\xymatrix{
& & & c_{10} \ar[dl]_{\zeta} \ar[d] & A_1 \ar[l]\\
& & c_{21} \ar[dl]_{\zeta/2} & c_{20} \ar[dl]_{\zeta} \ar[d] & A_2 \ar[l] \\
& c_{32} \ar[dl]_{\zeta/3}  & c_{31} \ar[dl]_{\zeta/2} & c_{30} \ar[dl]_{\zeta} \ar[d] & A_3 \ar[l]
\\
\idots&\idots&\idots&\vdots&\\}
\end{equation*}
For example, we calculate several first polynomials,
\begin{eqnarray}
&&P_{0}(t)\equiv 0,\qquad
P_1(t)=\cc{10}{A_1}, \\
&&P_2(t)=\cc{21}{\zeta A_1} (t-2T)+\cc{20}{A_1\exp(-2\gamma T)+A_2},\nonumber
\\
&&P_3(t)=\cc{32}{\dfrac{\zeta^2}{2} A_1} (t-3T)^2+\cc{31}{\zeta\big[A_1\exp(-2\gamma T)+A_2\big]}(t-3T)\nonumber
\notag \\
&&{}+ \cc{30}{\big\{\big[A_1\exp(-2\gamma T)+A_2\big]+\zeta T A_1\big\} \exp(-2\gamma T)+A_3}.\nonumber
\end{eqnarray}

\section{Solution of equation~\eqref{forK-step} for the processes with step-wise memory}\label{StepMemA}
\textbf{Solution for $K_0(t)$.} Using the parity condition~\eqref{even}, $K(t-\tau)=K(\tau-t)$, we split the region of integration in Eq.~\eqref{forK-step} into two parts, $0<\tau<t$ and $t<\tau<T$, and change the variables,
\begin{eqnarray}
&&\int_0^TK(t-\tau)d\tau =\int_0^tK(t-\tau)d\tau \\ && +\int_t^TK(\tau-t)d\tau\nonumber =\int_0^t K(\tau)d\tau+\int_0^{T-t}K(\tau)d\tau.\nonumber
\end{eqnarray}

Then differentiating Eq.~\eqref{forK-step} over $t$ we arrive at the delay differential equation,
\begin{eqnarray}
\label{K0-int-to-delayed}
K''(t)+2\gamma K'(t)-\xi K(t)=\xi K(T-t),
\,\,\,
t>0.
\end{eqnarray}

Applying operator $d^2/dt^2-2\gamma d/dt-\xi$ to Eq.~\eqref{K0-int-to-delayed}, we get the differential equation with constant coefficients,
\begin{eqnarray}
K_0''''(t)-(4\gamma^2+2\xi) K_0''(t)=0.
\end{eqnarray}

Its solution is
\begin{eqnarray}
&&K_0(t)=C_+ \exp(\eta t)+C_- \exp(-\eta t)+C_0+C_1 t,\nonumber\\[6pt]
&&\eta =\sqrt{4\gamma^2+2\xi}>0.
\end{eqnarray}
Substituting this solution into Eq.~\eqref{forK-step}, equating the coefficients at $t^0$, $t^1$ and $\exp(\pm\eta t)$, Eq.~\eqref{StepMemCF-0} is obtained after some simplification.

\textbf{General solution by iterating procedure.} Let us denote $K(t)=K_n(t)$ for the interval $nT<t<(n+1)T$. Then Eq.~\eqref{forK-step} transforms in a sort of recurrence relation for functions~$K_n(t)$,
\begin{eqnarray}
\label{Kn-intdiffeq}
&&K_n'(t)+2\gamma K_n(t)=\xi\Big[\int_0^{t-n T} K_{n}(t-\tau)d\tau\\
&&+\int_{t-n T}^T K_{n-1}(t-\tau)d\tau],
\,\,\,\,0 < n.\nonumber
\end{eqnarray}
Changing variables of integrations and differentiating over $t$ we obtain the differential equation with time shift (compare to Eq.~\eqref{K0-int-to-delayed}),
\begin{eqnarray}
\label{Kn-step-diffeq}
K_n''(t)+2\gamma K_n'(t)-\xi K_n(t) 
=-\xi K_{n-1}(t-T).
\end{eqnarray}
This differential equation is subject to two boundary conditions,
\begin{eqnarray}\label{Kn-step-boundary}
&&K_n(nT)=K_{n-1}(nT),\\
&&K_n'(nT)=-2\gamma K_{n-1}(nT)+\xi\int_{0}^{T} K_{n-1}(nT-\tau)d\tau.\nonumber
\end{eqnarray}

Then the problem can be solved iteratively:

1. $K_0(t)$ is defined by Eq.~\eqref{StepMemCF-0}.

2. Find $K_1(t)$ from Eq.~\eqref{Kn-step-diffeq} in the form of superposition of constant and exponential functions $\exp(\pm \eta t)$ (particular solution sourced from the constant term and the hyperbolic sine in $K_0(t)$ in the rhs), and $\exp[-(\gamma\pm\kappa) t]$ (the general solution of homogeneous equation), with
\begin{eqnarray}
\kappa=\sqrt{\gamma^2+\xi}.
\end{eqnarray}

3. Find $K_2(t)$ from Eq.~\eqref{Kn-step-diffeq} in the form of superposition of a constant term and the exponential functions $\exp(\pm \eta t)$, $\exp[-(\gamma\pm\kappa) t]$ and $t\exp[-(\gamma\pm\kappa) t]$ (particular solution sourced from $\exp[-(\gamma\pm\kappa) t]$ in $K_1(t)$ in the rhs).

It can be seen from the procedure that the function~$K_n(t)$ can be presented in the following form,
\begin{eqnarray}
&&K_n(t)=B_n+C_n\sinh[\phi_n-\eta (t-nT)] \nonumber \\[6pt]
&&+P_n^+(t) \exp[-(\gamma+\kappa)(t-nT)]\\ [6pt]
&&+P_n^-(t)\exp[-(\gamma-\kappa) (t-nT)],\nonumber
\end{eqnarray}
where $B_n$, $C_n$ and $\phi_n$ are indefinite constants, while $P_n^+(t)$ and $P_n^-(t)$ are some polynomial functions of the $(n-1)$-th order.

Substituting the last expression into Eq.~\eqref{Kn-step-diffeq}, equating prefactors at $\exp(\pm \eta t)$, $\exp[-(\gamma\pm\kappa) t]$, and at the constant  terms, we obtain coefficients $B_n$, $C_n$ as well as the recurrence relations for $\phi_n$. Then the function $K_n(t)$ can be expressed in the following form,
\begin{eqnarray}
\label{Kn-express}
K_n(t)&=&\bar{K}_n(t) +P_n^+(t)\exp[-(\gamma+\kappa) (t-nT)]\nonumber \\ [6pt] &&+P_n^-(t)\exp[-(\gamma-\kappa) (t-nT)],
\end{eqnarray}

\begin{widetext}
\begin{eqnarray}
&&\bar{K}_n(t)=K(0)\dfrac{2 \xi  \cosh \phi_0+\eta  (2 \gamma -\xi  T) (-1)^n\sinh[\phi_n-\eta(t-nT)]}{2 \xi  \cosh \phi_0+\eta (2 \gamma -\xi  T)  \sinh \phi_0},
\end{eqnarray}
\end{widetext}
\begin{eqnarray}
\phi_n&=&\phi_0+n\arctanh\frac{2 \gamma  \eta }{\eta ^2-\xi }.
\end{eqnarray}

\textbf{Iterative scheme for the polynomials~$P_n^\pm(t)$.} Substituting Eq.~\eqref{Kn-express} for $K_n(t)$ into Eq.~\eqref{Kn-intdiffeq}, we get the recurrence relations for $P_n^+(t)$ and $P_n^-(t)$,
\begin{eqnarray}
\label{Pn-step}
{P^\pm_n}''(t)\mp 2\kappa {P^\pm_n}'(t)=-\xi P^\pm_{n-1}(t-T).
\end{eqnarray}
It should be emphasized that the general solutions of Eq.~\eqref{Pn-step}, except polynomial summands, contain exponential terms with $\exp(\pm2\kappa t)$. Such terms  should be omitted. Therefore, two differential recurrence relations~\eqref{Pn-step} (with superscripts $\pm$) should be supplied with only two rather cumbersome boundary conditions that follow from Eqs.~\eqref{Kn-step-boundary},
\begin{widetext}
\begin{eqnarray}
&&P_n^+(nT)+P_n^-(nT)=
P_{n-1}^+(nT){\rm e}^{-(\gamma+\kappa) T}+P_{n-1}^-(nT){\rm e}^{-(\gamma-\kappa) T}+A_n,
\nonumber\\[6pt]
&&(\gamma-\kappa)P_n^+(nT)+(\gamma+\kappa)P_n^-(nT)=A_n'
-{P_n^+}'(nT)-{P_n^-}'(nT)
\nonumber\\[6pt]
&&+\xi\int_{0}^{T} \Big\{P_{n-1}^+(nT-\tau){\rm e}^{-(\gamma+\kappa) (T-\tau)}+P_{n-1}^-(nT-\tau){\rm e}
^{-(\gamma-\kappa) (T-\tau)}
\Big\}d\tau,
\\
&&A_n= \big[\bar{K}_{n-1}(nT)-\bar{K}_n(nT)\big],
\nonumber\\[6pt]
&&A_n'= -\bar{K}_n(nT)-2\gamma\bar{K}_n'(nT)+\xi\int_{0}^{T}\bar{K}_n(nT-\tau)d\tau.
\end{eqnarray}
\end{widetext}
We look for its explicit polynomial form as
\begin{eqnarray}
P_{n}^{\pm }(t)=\sum_{m=0}^{n-1} c_{nm}^{\pm }(t-nT)^m.\nonumber
\end{eqnarray}
According to Eq.~\eqref{Pn-step},
\begin{eqnarray}
&&c_{n(m+2)}^{\pm }(m+2)(m+1)\mp 2\kappa c_{n(m+1)}^{\pm }(m+1)\nonumber\\[6pt]
&&=-\xi c_{(n-1)m}^{\pm },
\quad
0\leqslant m\leqslant n-3,
\nonumber\\[6pt]
&&\mp 2\kappa c_{n(n-1)}^{\pm }(n-1)=-\xi c_{(n-1)(n-2)}^{\pm }.\nonumber
\end{eqnarray}
The second relation here can be reduced to
\begin{eqnarray}
c_{n(n-1)}^{\pm }=\Big(\pm\dfrac{\xi}{2\kappa}\Big)^{n-1}\dfrac{c_{10}^{\pm }}{(n-1)!}.\nonumber
\end{eqnarray}
Therefore, each coefficient $c_{n(m>0)}^\pm$ can be expressed via $c_{(n'<n)0}^{\pm}$. In turn, the coefficients $c_{n'0}^{\pm}$ can be found iteratively from the boundary condition,
\begin{widetext}
\begin{eqnarray}
&&c_{n0}^++c_{n0}^-=
A_n+\sum_{m=0}^{n-2} T^m\big[c_{(n-1)m}^{+ } {\rm e}^{-(\gamma+\kappa) T}+c_{(n-1)m}^{-}{\rm e}^{-(\gamma-\kappa) T}\big],
\nonumber\\[6pt]
&&(\gamma-\kappa)c_{n0}^++(\gamma+\kappa)c_{n0}^-=A_n'
-(c_{n1}^++c_{n1}^-)
\nonumber\\
&&+
\xi\sum_{m=0}^{n-2}\int_{0}^{T} \tau^m\big[c_{(n-1)m}^{+ }{\rm e}^{-(\gamma+\kappa) \tau}+c_{(n-1)m}^{-}{\rm e}
^{-(\gamma-\kappa) \tau}
\big]d\tau.\nonumber
\end{eqnarray}
\end{widetext}
The diagram below illustrates the scheme of the described calculations.
\begin{equation*}
    \xymatrix{
        & & & c_{10}^+ \ar[dl] \ar[d] \ar@/^/[r] &
        c_{10}^- \ar@/^/[l] \ar[dr] \ar[d] & & & \\
        & & c_{21}^+ \ar[dl] \ar[r] & c_{20}^+ \ar[dl] \ar[d] \ar@/^/[r] &
        c_{20}^- \ar@/^/[l] \ar[dr] \ar[d] & c_{21}^- \ar[dr] \ar[l] & & \\
        & c_{32}^+ \ar[dl] \ar[r]  & c_{31}^+ \ar[dl] \ar[r] & c_{30}^+ \ar[dl] \ar[d] \ar@/^/[r]  &
        c_{30}^- \ar@/^/[l] \ar[dr] \ar[d] & c_{31}^- \ar[dr] \ar[l] &  c_{32}^- \ar[dr] \ar[l] &
        \\
        \idots&\idots&\idots&\vdots&
        \vdots & \ddots & \ddots & \ddots \\}
\end{equation*}
Here we present the first instances of the recurrence relations that allow to calculate successively the coefficients $c_{10}^\pm$, $c_{21}^\pm$, $c_{20}^\pm$, etc.,
\begin{eqnarray}
&&A_1 = c_{10}^+ + c_{10}^-, \,\,\,A_1' = (\gamma-\kappa)c_{10}^+ + (\gamma+\kappa)c_{10}^-, \nonumber \\[6pt]
&&c_{21}^{\pm } = \pm\dfrac{\xi}{2\kappa}c_{10}^{\pm }, \,\, \nonumber\\[6pt]
&&c_{20}^+ + c_{20}^-=
A_2+\big[c_{10}^{+ } {\rm e}^{-(\gamma+\kappa)\nonumber T}+c_{10}^{-}{\rm e}^{-(\gamma-\kappa) T}\big], \nonumber
\end{eqnarray}
\begin{eqnarray}
&&(\gamma-\kappa)c_{20}^+ +(\gamma+\kappa)c_{20}^- =A_2'-(c_{21}^++c_{21}^-)\nonumber \\ [6pt]
&&+\xi\big[c_{10}^{+ }\dfrac{1-{\rm e}^{-(\gamma+\kappa) T}}{\gamma+\kappa}+c_{10}^{-}\dfrac{1-{\rm e}^{-(\gamma-\kappa) T}}{\gamma-\kappa}
\big], \nonumber
\end{eqnarray}
\begin{eqnarray}
c_{32}^{\pm }
&=&
\pm\dfrac{\xi}{2\kappa}c_{21}^{\pm }=\Big(\dfrac{\xi}{2\kappa}\Big)^2 c_{10}^{\pm },
\nonumber\\[6pt]
c_{31}^{\pm }
&=&
\dfrac{\xi c_{20}^{\pm }+2c_{32}^{\pm }}{\pm 2\kappa }=\dfrac{(2\kappa)^2\xi c_{20}^{\pm }+2\xi^2 c_{10}^{\pm }}{\pm (2\kappa)^3 },\nonumber
\end{eqnarray}
\begin{widetext}
\begin{eqnarray}
&&c_{30}^+ + c_{30}^-=
A_3+\big[(c_{20}^{+ }+c_{21}^{+ }T) {\rm e}^{-(\gamma+\kappa) T}+(c_{20}^{-}+c_{21}^{-}T){\rm e}^{-(\gamma-\kappa) T}\big],
\nonumber\\[6pt]
&&(\gamma-\kappa)c_{30}^+ +(\gamma+\kappa)c_{30}^-=A_3'
{}-(c_{31}^++c_{31}^-)
+\xi\Big\{c_{20}^{+ }\dfrac{1-{\rm e}^{-(\gamma+\kappa) T}}{\gamma+\kappa}+c_{20}^{-}\dfrac{1-{\rm e}^{[-(\gamma-\kappa) T]}}{\gamma-\kappa}
\notag\\[6pt]
&&+{}c_{21}^{+ }{}\dfrac{1-{\rm e}^{-(\gamma+\kappa) T}[1+T(\gamma+\kappa)]}{(\gamma+\kappa)^2}\!+\!c_{21}^{-}\dfrac{1-{\rm e}^{-(\gamma-\kappa) T}[1+T(\gamma-\kappa)]}{(\gamma-\kappa)^2}\Big\}.\nonumber
\end{eqnarray}
\end{widetext}

\end{document}